# Valuing Evaluation: Methodologies to Bridge Research and Practice

Petteri Kaskenpalo and Stephen G. MacDonell

*SERL, School of Computing and Mathematical Sciences*
*AUT University, Private Bag 92006*
*Auckland 1142, New Zealand*
*+64 9 921 9073*
*petteri.kaskenpalo@aut.ac.nz, stephen.macdonell@aut.ac.nz*

## ABSTRACT

*The potential disconnect between research and practice in software engineering (SE) means that the uptake of research outcomes has at times been limited. In this paper we seek to identify research approaches that are rigorous in terms of method but that are also relevant to software engineering practitioners. After considering the correspondence of several approaches to software systems research and practice we recommend a framework for applying grounded theory in SE research, as a means of delivering both robust and useful outcomes.*

**Keywords:** Research; practice; design science; software engineering; grounded theory.

## 1. INTRODUCTION

The desire to improve the relevance of software engineering (SE) research has been long evident. For instance, several calls for increased experimentation in SE research have been made; the growing use of systematic literature reviews is motivated in part by the goal of practice improvement; papers that traverse the literature and reflect on (among other things) the breadth and accessibility of the research methods utilized have appeared; and as recently as 2011 a special issue promoting the use of qualitative methods in empirical SE, due in part to their greater relevance to practice, was published in *Empirical Software Engineering*.

The focus in this paper is on research centered on the development and evaluation of SE artifacts. The design science research process (DSRP) [16] proposes an artifact-centred research methodology with the key steps of problem identification and motivation, objectives of a solution, design and development, demonstration, evaluation, and communication. At a glance, this process is analogous to the outline of many software development methods and software systems projects.

How does this process then account as a model for scientific research, and, considering the number of failed software projects that follow similar well-established development models, how might this process provide more relevant and trustworthy results?

In this paper we explore how the DSRP is proposed to achieve these goals of rigor and relevance, and then suggest a framework of using grounded theory (GT) [5] as a means of further increasing the value of artifact-based research. We discuss the key differences of developing software as a research artifact as opposed to doing so for a typical commercial goal, and by linking these two reasons for developing software, we explore opportunities for improving the relevance of research results to software practitioners.

We map several different life-cycle models, including those proposed previously for DS-based research along with well-established software and system models, and combine them into a reference model that enables us to discuss the steps of these processes together. We call this the Software Artifact-centred Research reference Process (SARP). Given constraints on space we focus in this paper on two steps in this process – problem identification and evaluation. For each step we discuss the requirements outlined in the literature for design science-based research, we outline the key activities that would happen in a non-research oriented software development process, we discuss the relevant GT processes and their benefits from both research and practice points of view, and in this we way establish a traceable trail of relevance throughout the process.

In the next section we review and compare some proposed DS research processes, we outline our reference model, we discuss the relationship, roles and contribution of the different academic computing disciplines and their matching professional roles in this model and in building theory. We then outline key GT principles and processes, followed by a discussion of how grounded theory can support the problem identification and evaluation steps of the SARP framework. Finally we conclude with a summary, and we outline intended future work.

## 2. DESIGN SCIENCE RESEARCH

The design science research process has been outlined by Peffers *et al.* [16] as a conceptual process and a mental model for the production and presentation of design science research in information systems (IS). Their model provides a roadmap for an artifact-centred research process and is consistent with earlier DS literature [1, 10, 14, 19] as well as the software engineering research methodology (SERM) proposed by Gregg *et al.* [8].

While the terminology used in these methodologies differs – for example, the DSRP includes steps of problem identification and motivation, objectives of a solution, design and development, demonstration, evaluation, and communication whereas the SERM comprises the phases of conceptualization, formalization and development – on closer inspection both methodologies are very similar and include similar expectations and goals.

The above cited authors also have a slightly different view of the nature of the involved artifact. The DSRP is said to represent a suitable methodology for research around *any* type of system-related artifact. Nunamaker *et al.* [14] include systems development as a key component of the process. Walls *et al.* [19] define Information Systems Design Theory and its products as the artifact, but also expect an actual system to be developed as a means of validation. The SERM process is suggested to involve an artifact as a product of a tangible nature, which takes the form of a software prototype or a formal specification of a solution. While it seems clear that a formal model of a solution would likely be sufficient for proving the validity of new concepts and theories to other researchers, it would be more difficult to use this as a means of communicating outcomes to practitioners, and impossible to “observe and measure how well the artifact supports the solution to the identified problem” [16] in a practical situation. However it has long been held that systems research *should* study the application of information technology (IT) in society [11], so it is crucial to assess such research in a practical setting (and preferably more than one), and to present outcomes in ways that are accessible and beneficial to practitioners [13]. For this reason, in addition to artifact-as-research views, we have considered the DSDM and V-model software development processes and the ITIL application management process [15] in the SARP framework (see Figure 1).

While the DSRP and SERM methodologies suggest a process of performing research based on developing a solution and then evaluating its suitability for addressing the identified problem, they also outline the need for this solution to be grounded in theory and to extend current knowledge boundaries [8]. However, these methodologies do not explicitly assist a researcher in doing this. The need for building a theory as part of computing research methodologies has been discussed in the literature [1, 9, 18]. Overall this should add rigor to research and assist in the evaluation of results in a broader context, ensuring a contribution to more general theories in the discipline. While this is often discussed in the context of information systems research, this need can also be identified in DS-based research in software engineering [6] and computer science [17].

We accept both the need for theory building and having relevance to practice as fundamental requirements for a rigorous and useful research methodology in SE. In this paper, we focus on a process that involves software development and a software prototype artifact, and we suggest a grounded theory (GT)-based approach in conducting software artifact-based research. We describe how this can integrate with SARP and how it should enable the results to be utilised by practising professionals.

## 3. ROLE OF GROUNDED THEORY

The principle of using Grounded Theory with the DS research process has been introduced by Adams and Courtney [1] as a theory-building technique, as a part of a multimethodical research framework integrating DS, GT and action research (AR), and by Fernandez and Lehmann [4] as a way to improve the rigor and relevance in information systems studies when investigating the role of such systems in organisational change via case studies.

Before we move on to consider the use of Grounded Theory with the DSRP, we briefly describe some of the key principles of the GT methodology and point the reader to material that provides a summary of the concepts and processes. Grounded theory has been discussed in numerous articles and textbooks in detail and to different degrees of clarity. Glaser [5, para.7] reminds us “that classic GT is simply a set of integrated conceptual hypotheses systematically generated to produce an inductive theory about a substantive area”, and continues [5, para.41]:

> “GT is not findings, not accurate facts and not description. It is just straightforward conceptualization integrated into theory - a set of plausible, grounded hypotheses. It is just that - no more - and it is readily modifiable as new data come from whatever source - literature, new data, collegial comments, etc. The constant comparative method weaves the new data into the sub-conceptualization.”

and summarizes the goal of GT being a “conceptual theory abstract of time, place and people” [5, para.42].

The use of grounded theory as an overarching methodology to study data from multiple sources helps in building a consistent theory as well as to moderate any bias the researcher's personal experience may introduce into the study by building the theory about the domain and via documented observations and developing conceptual coding of the analysed data. GT can use *any* data for building the conceptual theory. Evidence sources such as those suggested by Yin [20] for case study-based research are all suitable: documentation, archival records, interviews, direct observation, participant observation and physical artifacts. Sources can also include research literature reviews and theoretical or practical exploratory scenarios, software artifacts, system log-files, and system test records.

The GT process starts with open coding where the data is coded in order to identify substantive codes. This stage is literally *open*, as all available data is analysed. The aim is to relate coded incidents in the data to patterns that can be identified as categories and relationships between them. If these model a pattern, then new incidents in the data will fit these, and as the theory matures new incidents fit the existing categories better and better. Through the constant comparative method, coded incidents are constantly compared with other incidents to establish similarities in and differences between the incidents. The similarities support the identified categories, and new categories or properties for the existing categories are generated in order to better model the variation of incidents and their conditions. "Open coding allows the analyst to see the direction in which to take the study by theoretical sampling before he/she has become selective and focused on a particular problem. Thus, when he/she does begin to focus, he/she is sure of relevance." [5, para.49]

In grounded theory, data sources are chosen as they are needed rather than before the research begins, and it is an ongoing part of the research process to select suitable data sources according to the developing categories and the emerging theory. This is referred to as *theoretical sampling*. This approach is also in line with the three reasons suggested by Benbasat *et al.* for a case-based research strategy for studying such systems:

> "First, the researcher can study information systems in a natural setting, learn about the state of the art, and generate theories from practice. Second, the case method allows the researcher to answer "how" and "why" questions, that is, to understand the nature and complexity of the processes taking place. Questions such as, "How does a manager effectively introduce new information technologies?" are critical ones for researchers to pursue. Third, a case approach is an appropriate way to research an area in which few previous studies have been carried out." [2]

For the remainder of this paper, we use the definitions of the GT procedures as presented in [5], and refer to these as we discuss two of the DSRP steps.

# 4. THE FRAMEWORK

The reviewed DS research processes suggest clear steps for performing artifact-based research, but provide limited coverage of the actual methods that could be used during the process. As each of the stages produces more information from different viewpoints about the topic of the research, this information should be analysed and managed in a structured manner. Classic Grounded Theory [5] provides a suitable methodology to accompany the SARP steps for this purpose. We contend that a researcher is thus provided with a rigorous process and procedures, as the conceptual theory generation around the concern of the study, including the emergence of any problems that are to be addressed by the solution, is combined with the steps of practical implementation of the discovered/designed solution and its communication and evaluation. An overview of this integration is depicted in Figure 2.

This framework differs from, for example, the design science research framework presented by March and Smith [12] in that their framework consists of building, evaluating, theorizing and justifying as research activities, and our methodology emphasises the requirement of hypothesis-building or theorizing *prior* to taking other steps. Our approach builds on the theorizing as a required step that guides all the other activities. This way a greater emphasis is directed towards the *reasons* why the technology is required as well as towards assessing the resulting artifact with the relevant evaluation criteria, including any social and non-technical motivations for the work.

In the following two sub-sections we consider in more detail the role of GT in each of the DSRP research steps of problem identification and evaluation.

## 4.1 Problem Identification

Where do we start? How do we find a problem? What is the research question? Most methodologies and frameworks fall short in answering these questions or providing support in terms of mechanisms for addressing this fundamental part of a research process. Peffers *et al.* [16, p.89] note the following:

> "Define the specific research problem and justify the value of a solution. Since the problem definition will be used to develop an effective artifactual solution, it may be useful to atomize the problem conceptually so that the solution can capture the problems complexity."

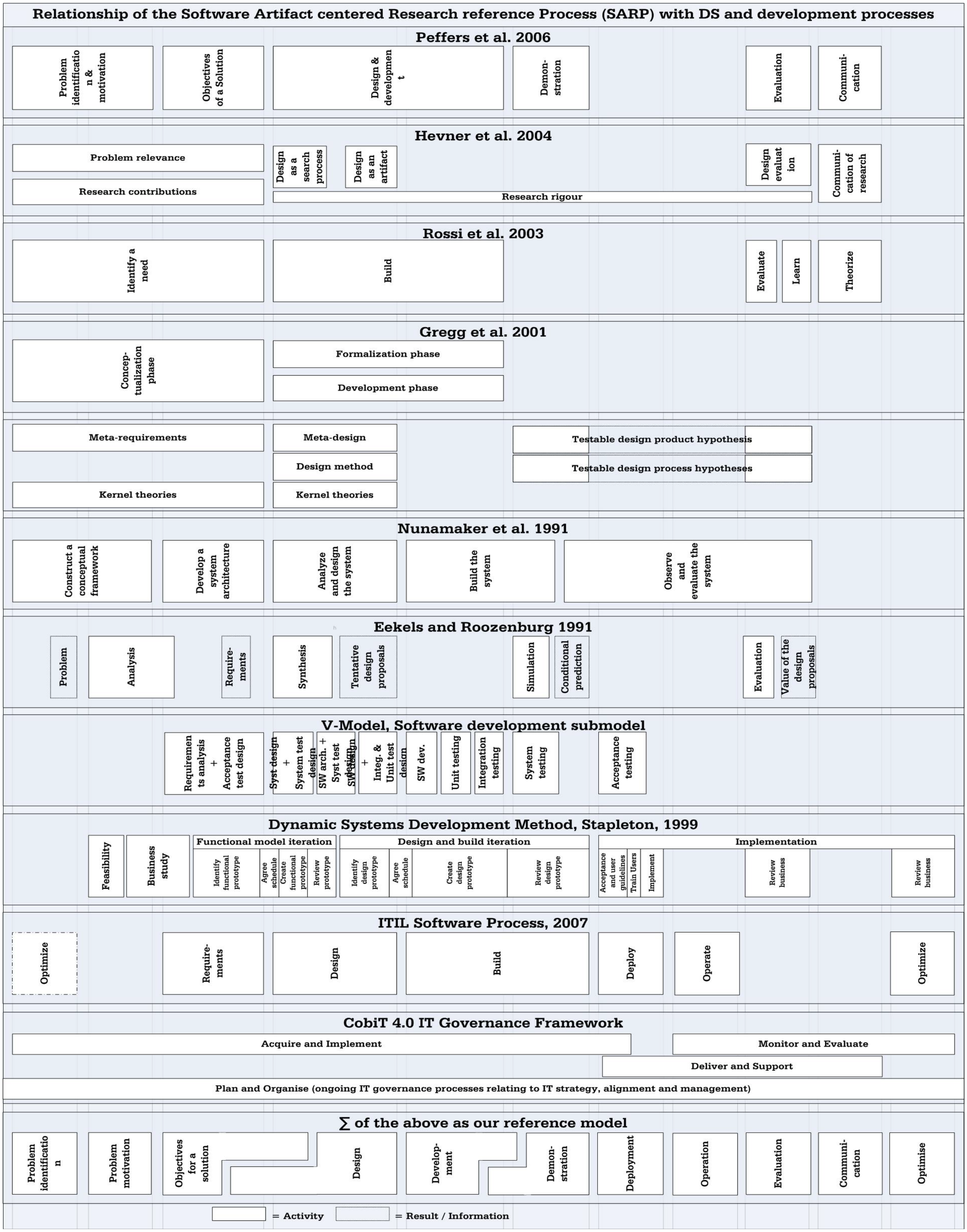

**Figure 1.** The Software Artifact-centered Research reference Process (SARP) as the aggregation of design science, software and systems processes

We argue that this complexity must be captured in order to provide grounding for the claim that the solution indeed addresses a valid problem (rigor), to understand the applicability of the solution to other similar problems (relevance), and to provide a grounded foundation for subsequent steps in the research process.

For this reason the problem identification and motivation stages are crucial. At this stage the focus is on developing a conceptual model that represents the domain under analysis and is grounded in the evidence available in the data. As concepts emerge, hypotheses are developed that explain their relationships with other concepts. This integration develops the grounded conceptual theory throughout the research process, and guides the selective data collection and the saturation of the theory.

In effect, design science-based research aims to satisfy two goals, those of research and those of design. Eekels and Roozenburg [3, p.200] summarize these two goals and their starting points:

- "the starting problem in the research cycle is a discrepancy between the facts and our set of truth-statements concerning these facts. The purpose of the process is adaption of the truth-statements (our knowledge) to the facts";
- "the starting problem in the design cycle is a discrepancy between the facts and our set of value-preferences concerning these facts. The purpose of the process is the adaption of the facts (through applications of the designed objects) to our value preferences";

in other words, the discovery of new knowledge through the research by examination of the real world, or the alteration of the real world, through the design, into one of the potentially realizable worlds. Which one of these starting points then should we adopt for a design science-based SE research project?

We suggest that either one is a potential starting candidate. Taking the research perspective, the researcher would start by identifying the problem and then explore solution candidates that could solve the problem. Alternatively, using the design view as a starting point, the researcher may suspect that a new technology, a new way of doing things, or the application of existing technology in a new situation could be beneficial and worth researching. With this approach the researcher is similarly challenged to discover real-world problems for which the system would provide a solution.

Either way, this is the beginning stage of applying GT in the process. The GT data analysis should be used immediately, and be applied to all available data with the focus on conceptualisation using the constant comparative method. As "GT stands alone as a conceptual theory generating methodology" [5, para.45], and can be described as a structured induction process, it is not only ideal for building a profound understanding of the domain and problem/opportunity under investigation, but also for developing a generalised theory. The analysis may well reveal several concerns, some more significant than others, but through conceptualisation a central concern will emerge.

This stage is conducted with open coding, and the researcher focuses on "patterns among incidents that yield codes and to rise conceptually above detailed description of incidents" [5, para.48]. Individual case studies may provide incidental evidence towards a pattern and suggest new categories, and additional case studies may provide incidental evidence that fit or reject the previous categories, and new categories may emerge.

Likely initial data sources include case studies, stakeholder interviews and the relevant literature. If the researcher is working in an area where field evidence does not exist or is not available, she or he can analyse the literature in order to identify weaknesses with other proposed, earlier solutions. In a design-driven approach, the researcher could use theoretical scenarios that would benefit from the solution and explore their relationship with the current reality. The researcher can describe this relationship and integrate any observations into the emerging theory.

Nunamaker *et al.* [14] state that in the beginning of the research process the researcher should construct a conceptual framework, which should address the stating of a meaningful research question. This step is further described as follows [14, p.635]: "Researchers should first justify the significance of research questions pursued. An ideal research problem is one that is new, creative, and important in the field." While we do not address several of the other stages here it is of note that both Nunamaker *et al.* [14] and Peffers *et al.* [16] also place significant emphasis on the objectives of the solution to be produced, which naturally form the necessary criteria for evaluation later in the process.

## 4.2 Evaluation

The main goal of the evaluation phase is to observe and measure how well the system solves the identified problems in terms of utility, quality and efficacy [16]. Hevner *et al.* [10] suggest a number of possible evaluation categories including observational, analytical, experimental, testing and descriptive methods. Following from the demonstration phase, the evaluation is based on data collected during and after the demonstration, deployment and operation phases, and focuses on analysing how well the objectives of the solution have been met, by examining simulation results and any experiences gained in configuring the system for laboratory scenarios, as well as analysing the data collected from test user groups.

Data collected during these phases is treated as new data for saturating the grounded theory, and as such it is coded and merged into the theory. This can include performance logs, system error logs, test user group interviews and so on. For example, error logs should be related to the user interview data in order to be able to eliminate user bias due to coding errors.

Note that we have included deployment and operational phases in the SARP framework in order to provide broader coverage of aspects that are relevant in the evaluation of software artifacts in practice, and to serve as a reminder of issues that need to be considered when conducting field experiments with the artifact.

The way to validate the results of course depends on the nature of the research, and an analysis of the objectives guides the evaluation process. If an artifact and the research process address several aspects, the validation process will in turn need to address each aspect, and significant planning is required to conduct experiments and data gathering in order to isolate the impact of each different aspect in the results. Eekels and Roozenburg [3, p.201] remind us of the difference between a scientific discovery and pure design: "The genuine scientist strives for explanation-power of his theories; the engineer is in most cases already satisfied with prediction-power." As researchers in an applied discipline such as software engineering, we should embrace both goals equally.

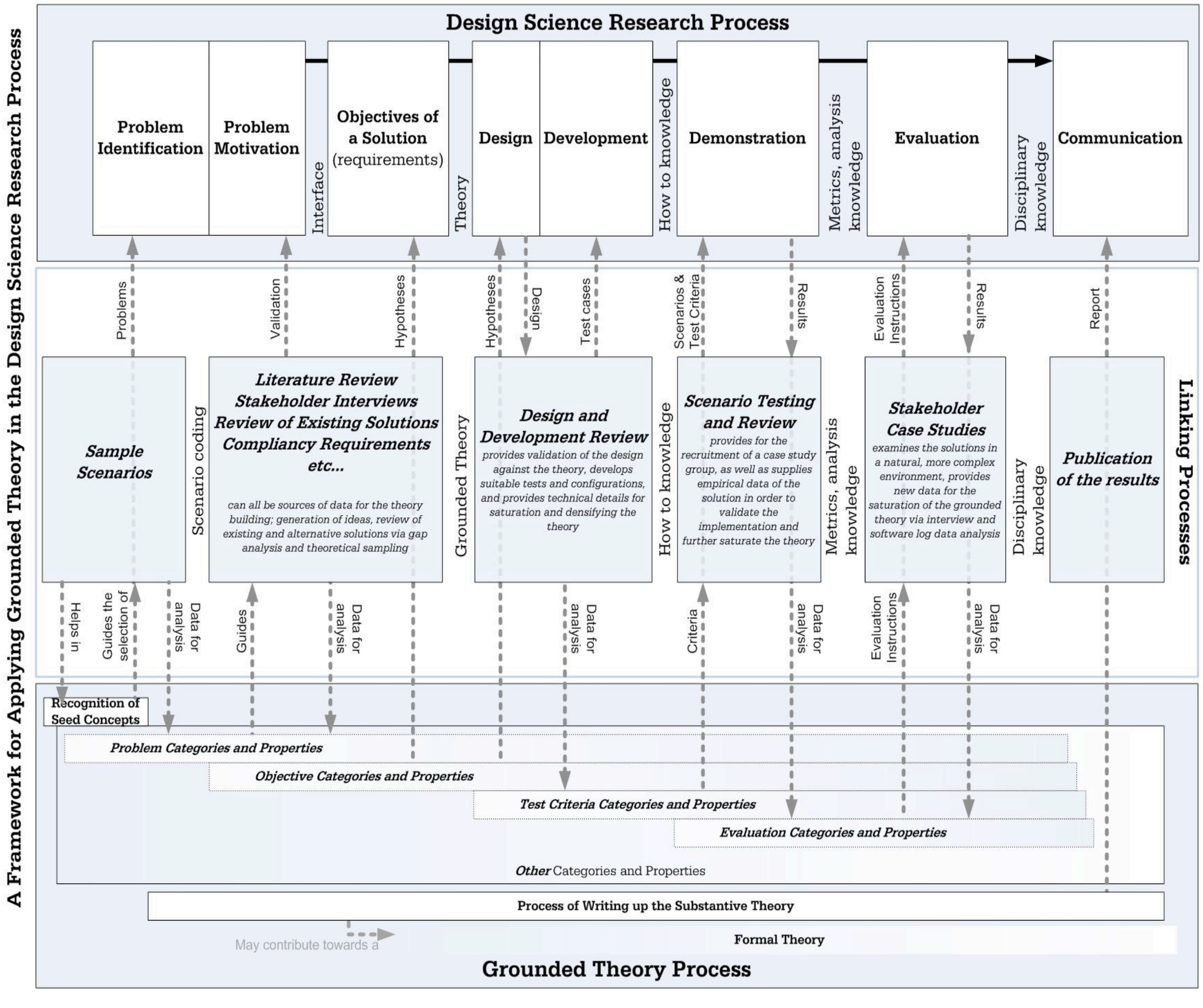

**Figure 2.** Development of Grounded Theory through project stages

# 5. SUMMARY

As a research community we have not always paid due attention to the robust evaluation of our work, sometimes relying on small-scale proofs of concept assessed with student subjects. While there are many valid reasons why such an approach has been predominant, it must contribute to the limited uptake of some research outcomes. In this paper we have introduced a framework that enables a design science research process to be carried out with the support of grounded theory in order to lead to relevant and rigorous research and practice outcomes arising from the development of software artifacts. We have briefly presented the entire approach in diagram form and have described the proposed conduct of two steps – problem identification and evaluation.

We are currently making extensive use of the design science research process in our work and are looking to test out the proposed framework by using GT as the means of identifying software system needs, informing development of appropriate solutions, and evaluating their utility, quality and efficacy in both testing and in actual use. We also intend to compare the characteristics and coverage of our approach to other evaluation schemes centered on artifacts [7, 13].